\input{psfig.sty}

\documentclass[preprint]{aastex}

\begin{document}
 
\title{Detection of a Hard Tail in the X-ray Spectrum of the Z Source GX~349+2}

\author{T. Di Salvo\altaffilmark{1,2},
N. R. Robba\altaffilmark{1}, R. Iaria\altaffilmark{1},
L. Stella\altaffilmark{3,4}, L. Burderi\altaffilmark{3}, 
G. L. Israel\altaffilmark{3,4}}
\altaffiltext{1}{Dipartimento di Scienze Fisiche ed Astronomiche, 
Universit\`a di Palermo, via Archirafi n.36, 90123 Palermo, Italy; 
disalvo@gifco.fisica.unipa.it.}
\altaffiltext{2}{Astronomical Institute "Anton Pannekoek," University of 
Amsterdam and Center for High-Energy Astrophysics,
Kruislaan 403, NL 1098 SJ Amsterdam, the Netherlands.}
\altaffiltext{3}{Osservatorio Astronomico di Roma, Via Frascati 33, 
00040 Monteporzio Catone (Roma), Italy; stella@coma.mporzio.astro.it.}
\altaffiltext{4}{Affiliated with the International Center for Relativistic 
Astrophysics.}
  
\begin{abstract}
We present the results of a BeppoSAX observation of the Z source GX~349+2
covering the energy range 0.1--200~keV. The presence of  
flares in the light curve indicates that the source was in the
flaring branch during the BeppoSAX observation. 
We accumulated energy spectra separately for the non-flaring intervals
and the flares. In both cases the continuum is well described by a 
soft blackbody ($k T_{\rm BB} \sim 0.5$~keV) and a Comptonized spectrum 
corresponding to an electron temperature of $k T_{\rm e} \sim 2.7$~keV, 
optical depth $\tau \sim 10$ (for a spherical geometry), and seed photon 
temperature of $k T_{\rm W} \sim 1$~keV.  All temperatures tend
to increase during the flares.
In the non-flaring emission a hard tail dominates the spectrum above
30~keV. This can be fit by a power law with photon index $\sim 2$,
contributing $\sim 2\%$ of the total source luminosity over the BeppoSAX
energy range.  
A comparison with hard tails detected in some soft states of black hole
binaries suggests that a similar mechanism could originate 
these components in black hole and neutron star systems. 
\end{abstract}

\keywords{accretion discs -- stars: individual: GX~349+2 --- stars: neutron 
stars --- X-ray: stars --- X-ray: spectrum --- X-ray: general}

\section{Introduction}

Low Mass X-ray Binaries (LMXB) are usually divided into Z and Atoll 
sources, according to the path they describe in a X-ray Color-Color 
Diagram (CD) or hardness-intensity diagram (Hasinger \& van der Klis 1989). 
The six known Z sources in the Galaxy are among the most luminous LMXBs
and close to the Eddington limit ($L_{\rm Edd}$) for a 1.4 $M_\odot$ 
neutron star (NS). 
The instantaneous position of an individual source in the CD is
an indicator of the mass accretion rate (e.g. Hasinger et al. 1990), 
which most likely increases along the Z track from the top left to the 
bottom right.

The LMXBs of the Atoll class usually have lower luminosities than Z sources
(generally in the range 0.01--0.1 $L_{\rm Edd}$). 
These sources can be found in soft or hard states (see Barret et al. 2000). 
In the hard state their spectrum is dominated at high energies by a power 
law, sometimes showing an exponential cutoff at energies of
$\ga 100$~keV (see e.g. Barret et al. 2000, and references therein).
On the other hand, the spectrum of the Z sources
is much softer, with cutoff energies 
usually well below 10~keV.  However, hard tails were occasionally
detected in their spectra. The first detection was in the spectrum 
of Sco X--1; beside the main X--ray component, at a temperature of 
$\sim 4$~keV, Peterson \& Jacobson (1966) found a hard component dominating 
the spectrum above 40~keV.  The latter component
was observed to vary as much as a factor of 3. Since then hard
tails were repeatedly searched for in the spectrum of Sco X--1 and other Z
sources: they were detected on occasions (e.g. Buselli et al. 1968; Riegler,
Boldt, \& Serlemitsos 1970; Agrawal et al. 1971; Haymes et al. 1972), but
in most cases they were not found, perhaps due to pronounced variations 
(e.g. Miyamoto \& Matsuoka 1977, and references therein; Soong \& 
Rothschild 1983; Jain et al. 1984; Ubertini et al. 1992). 
Recently a hard tail was detected in another Z source, GX~17+2,
observed by BeppoSAX in a broad energy range (0.1--200~keV). In this 
case the intensity variations of the hard tail were clearly correlated 
with the source spectral state: a factor of 20 decrease 
was observed moving from
the Horizontal Branch (HB) to the Normal Branch (NB; Di Salvo et al. 2000).
The presence of a variable hard tail in Sco X--1 was confirmed by
OSSE and RXTE observations
(Strickman \& Barret 2000; D'Amico et al. 2000). A hard tail was also 
detected in Cir X--1 (Iaria et al. 2001) and in type-II bursts from the 
Rapid Burster (Masetti et al. 2000). 

GX~349+2, also known as Sco X--2, is one of the six Z sources. 
Similar to the case of Sco X--1, GX~349+2 possesses a short and
underdeveloped HB (if at all).
The source variability in the frequency range below 100 Hz is closely
correlated with the source position on the X-ray CD, as in other Z sources.  
Twin quasi periodic oscillations at kHz frequencies (kHz QPO) were 
detected in the upper and middle NB (Zhang, Strohmayer, 
\& Swank 1998).
In this paper we report the results of a spectral study of GX~349+2 
in the energy range 0.1--200~keV based on data obtained with BeppoSAX.
This led to the detection of a hard component, fitted by a power law 
with photon index $\sim 1.8$ or by a thermal bremsstrahlung with 
$k T \sim 120$~keV.  

\section{Observations and Analysis}

The Narrow Field Instruments (NFI) on board BeppoSAX satellite 
are four co-aligned instruments which cover more
than three decades in energy, from 0.1~keV up to 200~keV, with good
spectral resolution over the whole range (see Boella et al. 1997a for detailed
description of BeppoSAX instruments). 
These are two Medium Energy Concentrator Spectrometers, MECS 
(position sensitive proportional counters operating in the 1.3--10~keV band,
Boella et al. 1997b), a Low Energy Concentrator Spectrometer, LECS 
(a thin window position sensitive proportional counter with extended low 
energy response, 0.1--10~keV; Parmar et al. 1997), a High Pressure Gas 
Scintillation Proportional Counter (HPGSPC; energy range of 7--60 keV; 
Manzo et al. 1997) and a Phoswich Detection System (PDS; energy range of 
13--200 keV; Frontera et al. 1997).

GX~349+2 was observed by BeppoSAX from 2000 March 10 20:42 UT to 
March 11 23:10 UT, 
with an effective exposure time of $\sim 16$ ks in the LECS, $\sim 45$ ks
in the MECS, $\sim 45$ ks in the HPGSPC, and $\sim 22$ ks in the PDS.  
We selected the data for scientific analysis 
in circular regions centered on the source with $8'$ and $4'$ radius for 
LECS and MECS, respectively. The background subtraction
was obtained with standard methods by using blank sky observations. 
The background subtraction for the high-energy (non-imaging) instruments 
was obtained by using off-source data for the PDS and Earth occultation data 
for the HPGSPC.   
There is no evidence indicating the presence of 
contaminating sources in the FOVs of the HPGSPC and PDS,
either in the on-source and off-source positions. In fact 
the off-source PDS count rates are in the expected range and 
the HPGSPC and PDS spectra align well with each other and the MECS
spectra.

Relative normalizations of the four NFIs 
were treated as free parameters in the model fitting, except for the MECS
normalization that was fixed to a value of 1. We checked after the fitting
procedure that these normalizations were in the standard range for each
instrument\footnote{See the BeppoSAX handbook at 
http://www.sdc.asi.it/software/index.html.}.
The energy ranges used in the spectral analysis were: 
0.12--4~keV for the LECS, 1.8--10~keV for the MECS, 8--40~keV for the HPGSPC,
and 15--200~keV for the PDS. We rebinned the energy spectra in order to
have approximately the same number of bins per instrument resolution element
across the entire energy range.  
Moreover a 1\% systematic error was applied to all these spectra in order 
to take into account calibration residuals.

In Figure 1 we show the 200~s binned MECS light curve of GX~349+2 in the 
1.8--10~keV range.  
A flare, lasting $\sim 4$ ks, is clearly apparent close to the 
beginning of the observation, followed by a series of smaller flares.
In Figure 2a we show the CD of GX~349+2,
where we defined the Hard Color (HC) as the ratio of the counts in the 
7--10.5~keV to the 4.5--7~keV energy bands, and the Soft Color (SC) as the 
ratio of the counts in the 4.5--7~keV to the 1.8--4.5~keV bands. 
The SC and HC relationship versus the source intensity in the 1.8--10.5~keV 
is also shown in Figure 2b (upper and lower panel, respectively).
The color variations in these diagrams are mainly due to the spectral changes
associated with the flares, during which both the HC and the SC increase.
In consideration of presence of fairly large flares
in the light curve and the shape of the source variations in 
the CD, we identify the state of the source with the flaring branch (FB) 
of the Z path.  Hints of the NB/FB vertex are visible in the diagrams
at $SC \sim 0.2$ and $HC \sim 0.3$.
We extracted energy spectra during the non-flaring intervals, corresponding
to SC lower than 0.48, and during the flares, corresponding to SC higher than 
0.48.
\begin{figure}[h!]
\centerline
{\psfig
{figure=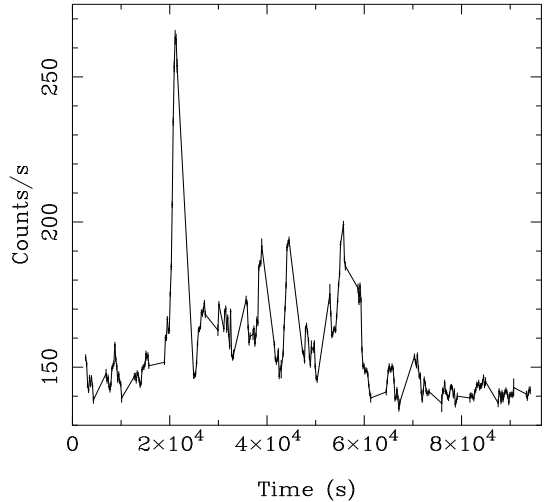,height=10.0cm,width=8.0cm}}
\caption{Light curve of GX~349+2 in the 1.8--10~keV (MECS data).
Each bin corresponds to 200 s integration time. }
\label{fig1}
\end{figure}
\begin{figure}[h!]
$$\psfig
{figure=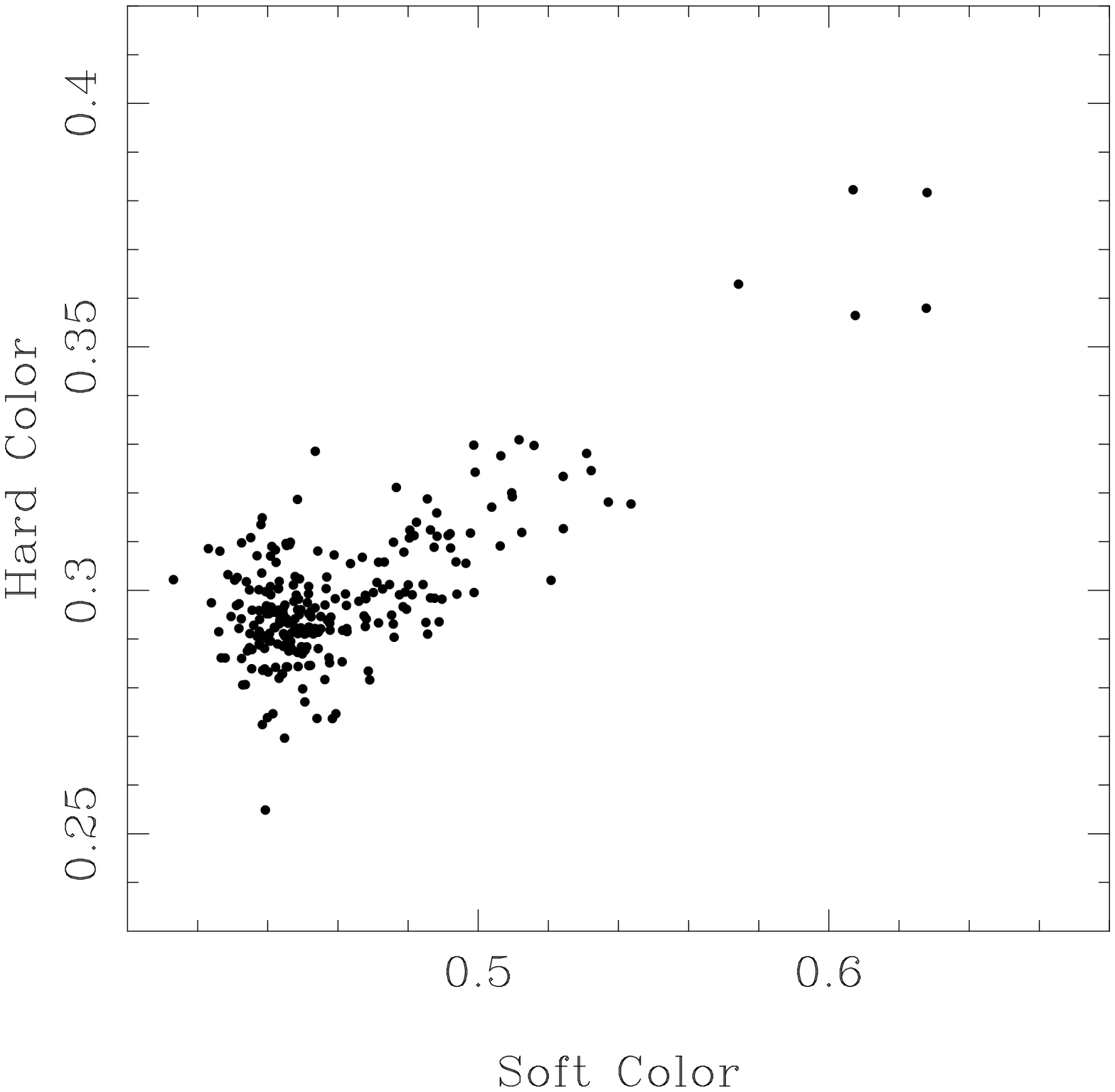,height=10.0cm,width=8.0cm}\qquad
\psfig{figure=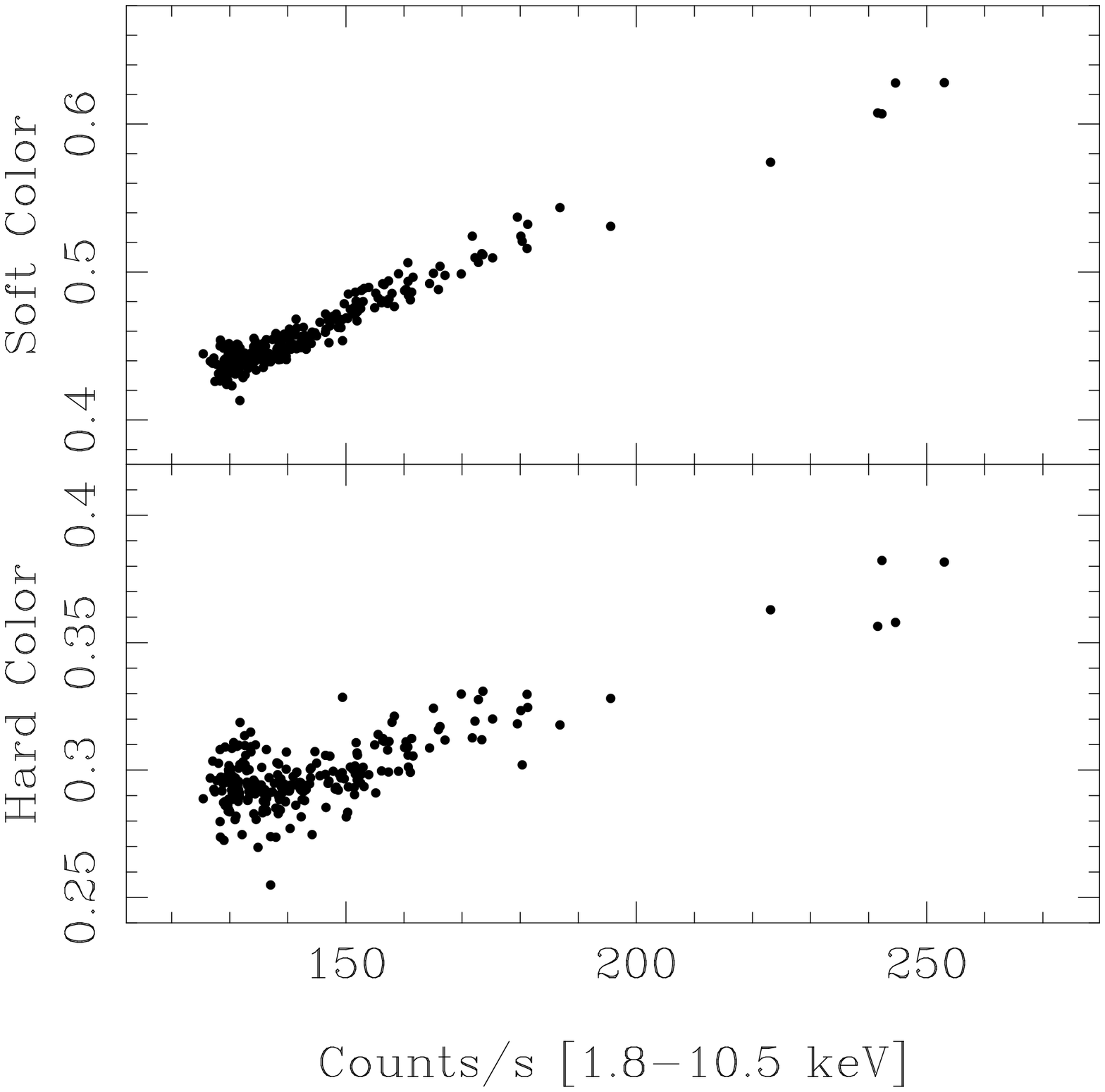,height=10.0cm,width=8.0cm}$$
\caption{a) Color-Color Diagram of GX~349+2.
The Hard Color is the ratio of the counts in the 7--10.5~keV and the 
4.5--7~keV energy bands, and the Soft Color is the ratio of the counts in  
the 4.5--7~keV and 1.8--4.5~keV energy bands. Each bin corresponds to 200 s.
b) Soft Color (upper panel) and Hard Color (lower panel)
versus the source count rate in the MECS (energy range 1.8--10.5~keV).
\label{fig2a} \label{fig2b}}
\end{figure}

We tried several two-component models to fit the source X-ray spectrum during 
the non-flaring state. In particular we tried a blackbody or a disk 
multicolor blackbody ({\tt diskbb} in XSPEC, Mitsuda et al. 1984) to describe 
the softer component, and blackbody or Comptonization 
models such as a power law with high energy exponential cutoff, {\tt compst} 
(Sunyaev \& Titarchuk 1980), {\tt comptt} (Titarchuk, 1994),
and {\tt pexriv} ({\it i.e.} a power law with exponential cutoff with its
reflection component, Magdziarz \& Zdziarski 1995) to describe the 
harder component. None of these models could well describe the source
X-ray spectrum in the whole BeppoSAX energy range, because an excess
of emission was present above $\sim 30$~keV (see Fig.~3a, upper and middle
panels).  We then considered the
low energy part (0.1--30~keV) of the spectrum, where 
we obtained the best fit by using a blackbody plus {\tt comptt} 
for the continuum:  
this model gave a reduced
chisquare of $\chi^2_{\rm r} = 1.19$, while we obtained  
$\chi^2_{\rm r}$ from 1.49 to 2.35 for the other models we fitted.
An emission line at
$\sim 6.7$~keV and an edge at $\sim 8.5$~keV, probably produced by highly 
ionized iron, are also necessary (probability of
chance improvement of the fit $< 10^{-20}$ and $\sim 8.6 \times 10^{-5}$
for the addition of an emission line and an absorption edge,
respectively).
The fit is further improved by the addition of a low energy Gaussian 
emission line at $\sim 1.16$~keV (chance probability of $\sim 7 \times 
10^{-9}$).
\begin{figure}[h!]
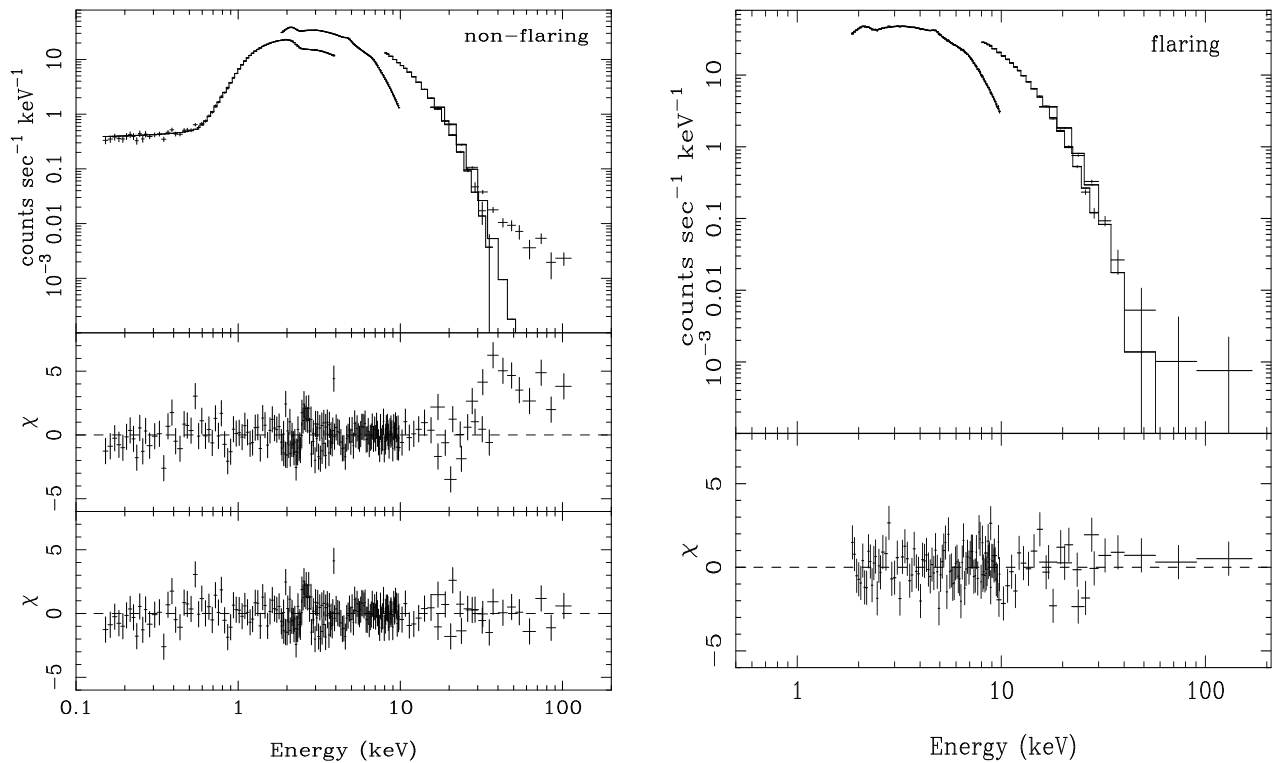

$$\psfig
{figure=figure_3a.ps,height=10.0cm,width=8.0cm}\qquad
\psfig{figure=figure_3b.ps,height=10.0cm,width=8.0cm}$$
\caption{a) Broad band spectrum of GX~349+2 during the non-flaring 
emission together with the best fit two-component continuum model 
(blackbody plus {\tt comptt}, upper panel), and the corresponding residuals 
in unit of $\sigma$ (middle panel). 
Residuals in unit of $\sigma$ (lower panel) with respect to the best fit 
model reported in Table~1, including a power-law hard component.
b) GX~349+2 spectrum during flaring intervals together with
the best fit model (upper panel), and residuals in unit of $\sigma$ 
(lower panel).  \label{fig3a} \label{fig3b} }
\end{figure}

A two-component continuum model was not sufficient to well fit the non-flaring
spectrum in the whole BeppoSAX range. 
A hard excess is clearly visible above 30~keV with respect to any of the
two-component models that we tried.  For example, the best fit
model described above, when fitted in the whole 0.1--200~keV range, gives 
rise to the residuals (in units of $\sigma$) shown in Figure 3a (middle panel).
A significant improvement of the fit was obtained by
adding to the model a power law with photon index $\sim 1.9$, which
eliminates the residuals at high energies (see Fig.~3b, lower panel). 
With the addition of this component the $\chi^2$ decreases from 397 
(for 191 d.o.f) to 220 (189 d.o.f.). An F-test indicates that the 
probability of chance improvement is negligibly small.
The contribution of this power law component to the unabsorbed 0.1-200~keV 
luminosity is $\sim 2\%$. 
Alternative models for the hard excess cannot be excluded. 
Using a thermal bremsstrahlung for the hard component we obtain a
similarly good fit for a temperature of $kT_{\rm TB} = 120 \pm 50$~keV. 
A high energy power-law spectrum might in principle be produced by 
Comptonization of seed photons with a blackbody distribution in a optically 
thin, hot region. Therefore we also tried to fit the hard component using 
the Comptonization model {\tt compbb} (Nishimura, Mitsuda, \& Itoh 1986), 
which, however, gave a worse fit. 
In particular we substituted the blackbody plus power law of the best fit 
model with {\tt compbb}, obtaining a $\chi^2_r = 1.41$ (for 189 d.o.f.)
and a temperature of the optically thin Comptonizing region higher than 
150 keV.
The high $\chi^2_r$ is due to large residuals (up to $4-5\ \sigma$) in the 
energy range above $\sim 70$~keV, because of a sharp cutoff in the model, 
which is not present in the data. In other words, because these data do not 
show a high energy cutoff in the BeppoSAX range, this implies a high 
electron temperature ($\ga 100$~keV) or a non-thermal origin. 

No useful LECS data were obtained during the flares, because of the low 
exposure time: due to UV contamination problems, the LECS is usually 
operated only at satellite night time, resulting in a much reduced on-source
time; therefore low energy data (0.1--1.8~keV) were not available in this case.
The spectrum during the flare cannot be described by a single component,
such as a blackbody or a {\tt comptt}. As in the case of the non-flaring 
spectrum, we obtained a good fit using a blackbody plus {\tt comptt} model.  
The addition of an iron line, with centroid energy and
width fixed to the best fit values found in the non-flaring spectrum,
gave an improvement of the fit at 99.9\% confidence level.  
On the other hand the addition of a power law at high energies did not 
improve the fit: the $\chi^2$ decreases from 145 
(for 119 d.o.f) to 144 (118 d.o.f.), which is clearly not significant. 
To study the variation of the power law intensity with respect to the 
non-flaring spectrum we fixed the photon index fixed to the best fit value 
of 1.9 found in the non-flaring spectrum,
and calculated the corresponding power-law normalization. 
We find an upper limit (at 90\% confidence level) on the power-law unabsorbed 
0.1--200 keV flux of $3.6 \times 10^{-10}$ ergs~cm$^{-2}$~s$^{-1}$ 
($\sim 1\%$ of the unabsorbed 0.1--200 keV flux during the flare).
Therefore we conclude that there is evidence that 
the hard component weakens during the flare, although the upper limit
is still compatible with the best fit power-law normalization during the
non-flaring emission.

We report in Table~1 the best fit parameters corresponding to the spectra 
during the non-flaring and the flaring intervals. These spectra are shown 
in Figure 3 (top panels, a and b respectively), 
together with the residuals with respect to the best fit models
(bottom panels).  The unfolded spectra are also shown in Figure~4 
(a and b, respectively), together with the components of the best fit
models reported in Table~1.
\begin{figure}[h!]
$$\psfig
{figure=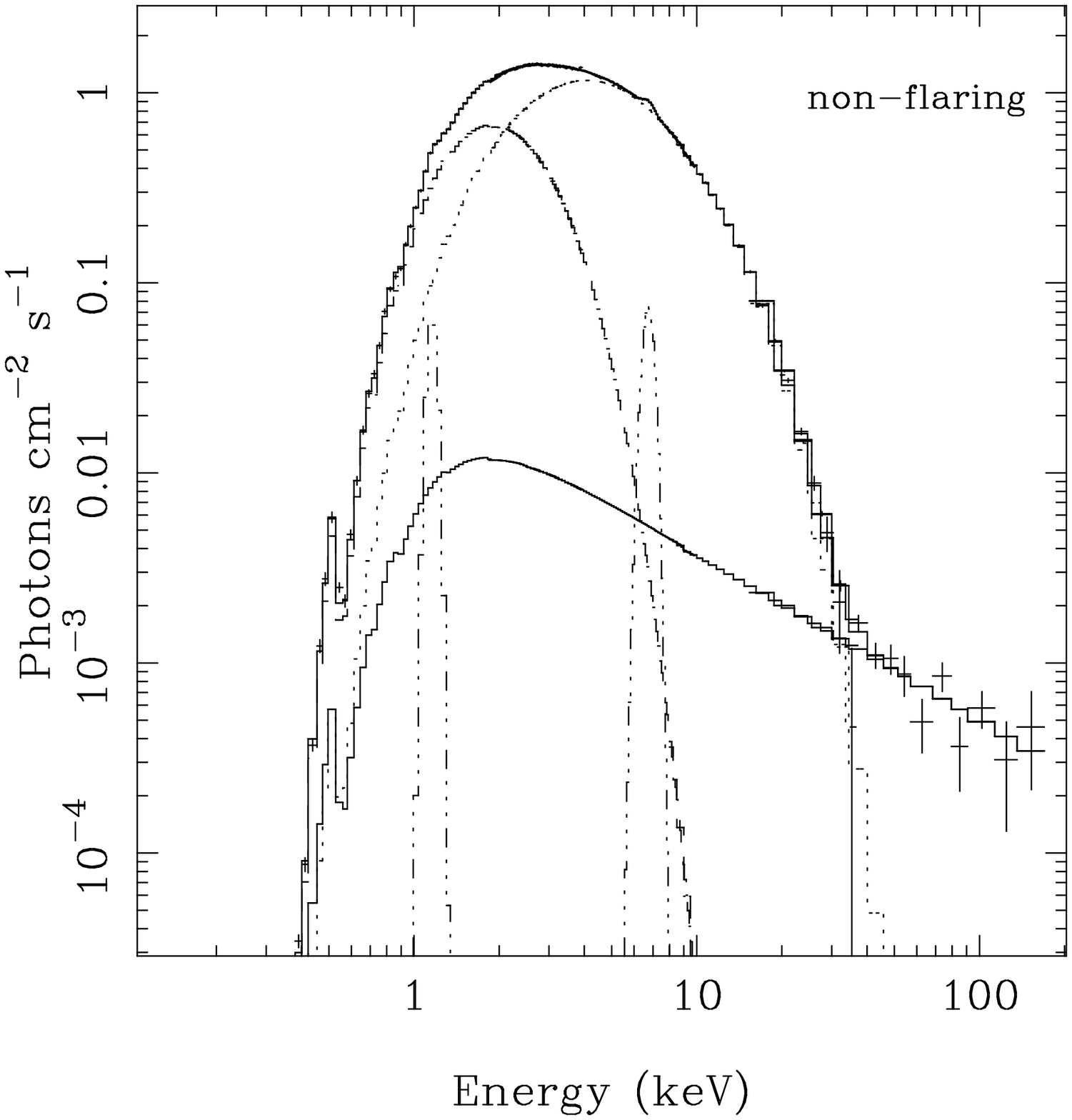,height=10.0cm,width=8.0cm}\qquad
\psfig{figure=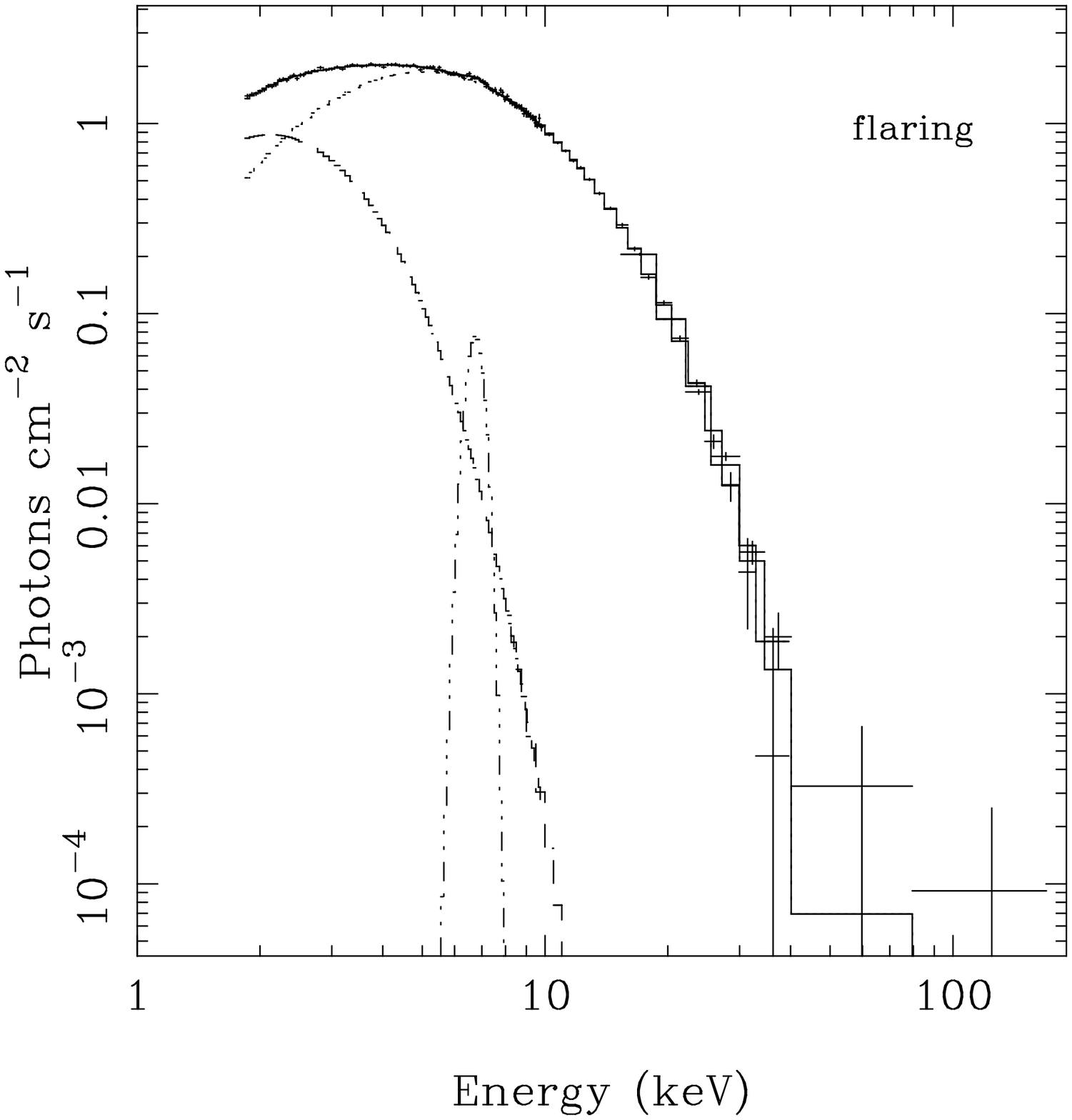,height=10.0cm,width=8.0cm}$$
\caption{a) Unfolded spectrum of the GX~349+2 non-flaring emission 
and the best fit model of Table 1, shown in this figure as the solid line 
on top of the data. The individual model components are also 
shown, namely the blackbody (dashed line), the Comptonized spectrum 
({\tt comptt} model, dotted line), two Gaussian emission lines at 
$\sim 1.2$ keV and $\sim 6.7$ keV (dot-dot-dot-dashed lines), and
the power-law (solid line).
b) Unfolded spectrum of the GX~349+2 flaring emission 
and the best fit model of Table 1, shown as the solid line 
on top of the data. The components of the model are also shown, namely
the blackbody (dashed line), the {\tt comptt} model,
(dotted line), and a Gaussian emission lines at $\sim 6.7$ keV 
(dot-dot-dot-dashed line).}
\label{fig4a}
\label{fig4b}
\end{figure}

\section{Discussion}

We fitted the BeppoSAX energy spectra of GX~349+2, extracted at different 
positions of the source in the CD. The best fit model up to energies of
$\sim 30$~keV consists of a blackbody and a Comptonization spectrum 
(described by the {\tt comptt} model), two emission lines and an absorption 
edge. 
The blackbody is at a temperature $kT_{\rm BB} \sim 0.5-0.6$~keV, and
the radius of the blackbody (spherical) emitting region
is $R_{\rm BB} \sim 35$~km (using a distance of 5 kpc, Cooke \& Ponman 1991;
Christian \& Swank 1997). 
The temperature of the soft seed photons for the Comptonization is 
$k T_{\rm W} \sim 1$~keV. 
These are Comptonized in a hotter ($k T_{\rm e} \sim 3$~keV) region of 
moderate optical depth ($\tau \sim 10-12$ for a spherical geometry).
The radius of the region emitting the seed-photon Wien spectrum, 
calculated as in In 't Zand et al. (1999), is 
$R_{\rm W} = 3 \times 10^4 D \sqrt{f_{bol}/(1+y)}/(kT_W)^2\; {\rm km} 
\sim 7-9$~km,
where $D$ is the distance in kpc, $f_{bol}$ is the bolometric flux in 
ergs cm$^{-2}$ s$^{-1}$, and $k T_W$ is in keV.
A broad ($\sim 0.7$~keV FWHM) iron K$\alpha$ emission line is present at 
$\sim 6.7$~keV, with equivalent width $\sim 30-70$ eV, accompanied by
an absorption edge at $\sim 8.5$~keV. The high energy of both the line and
edge indicates that these features are produced in a highly ionized region
(corresponding approximately to Fe XXV). 
We also found evidence for an emission line at $\sim 1.2$~keV, with equivalent 
width of $\sim 20$ eV, which can be associated with emission from the L-shell
of Fe~XXIV or, perhaps, the K-shell of Ne~X (see e.g. Kallman et al. 1996).  
We note that the main differences in the continuum below 30~keV between
the spectra during the flares and the non-flaring emission are in the 
temperatures of the blackbody and Comptonized components: 
both temperatures are higher during the flares than during 
the non-flaring emission, while the optical depth of the Comptonized
component is smaller during the flares.  The iron emission line  
remains approximately unchanged.  

A hard component is required to match the spectrum above 30~keV during the 
non-flaring emission. 
This component can be fit by a power-law with photon index $\sim 2$,
contributing $\sim 2\%$ of the 0.1--200~keV source flux. 
This component is not required 
in the flare spectrum, where it probably weakens  
contributing a smaller fraction of the source flux. 
Evidence for such a hard component was previously found in this source by 
the {\it Ariel V} satellite (Greenhill et al. 1979), and was fit by a 
thermal spectrum with $k T \sim 30$~keV. In our data this component 
is much harder, corresponding to a temperature of $\sim 120$~keV.
We note that the high energy instrument on board {\it Ariel V} had a 
field of view of $8^\circ$ FWHM, much larger than the PDS field of view.

The smaller PDS field of view,
$1.3^\circ$ FWHM, reduces the possibility of the presence of 
contaminating sources. The eclipsing binary system 4U~1700--37
(the only nearby X-ray source brighter than 1~$\mu$Jy, e.g. 
Valinia \& Marshall 1998),
is $1.4^{\circ}$ away from GX~349+2, and out of the PDS field of view.
Another possible contaminating source is the
hard diffuse emission of the Galactic ridge.  Using data from 
Valinia \& Marshall (1998), for latitudes
$1.5^{\circ}<|b|<4^{\circ}$ and longitudes $|l|<15^{\circ}$ 
(the region of GX~349+2) the flux of the diffuse Galactic emission is 
$\sim 3.2 \times 10^{-11}$ ergs cm$^{-2}$ s$^{-1}$ in the 10--60 keV
energy range for the effective solid angle of the PDS FOV.  
This is one order of magnitude 
lower than the flux of the hard power-law component we detected from 
GX~349+2   
in the same energy range.
Given the flux of the power-law component in the 13--80~keV energy range,
which is $\sim 1.2 \times 10^{-10}$~ergs cm$^{-2}$ s$^{-1}$, we  
calculated the probability to find an active galactic nucleus in the 
PDS FOV at a flux level equal or higher than $\sim 1.2 \times 
10^{-10}$~ergs cm$^{-2}$ s$^{-1}$.  We find a small probability of $\sim 2 
\times 10^{-3}$ (see Fiore \& Tamburelli 2000, in preparation; Levine et 
al. 1984). 
We can also exclude that this hard component is an instrumental
feature, because in other BeppoSAX observations of soft sources there was no
evidence of a hard excess. A clear example of this is given by the BeppoSAX
spectrum of GX~17+2 in the lower NB (Di Salvo et al. 2000).  We find that
a power law with photon index 1.9, like that detected in this paper, is 
incompatible with those data, which give a 90\% upper limit on the 
power-law unabsorbed 0.1--200 keV flux of 
$2.5 \times 10^{-11}$ ergs~cm$^{-2}$~s$^{-1}$.
It therefore is plausible that this hard component represents emission 
from GX~349+2.

A similar hard component has also been observed in other Z sources,
indicating that this is probably a common feature of these sources.
The presence (or strength) of these components   
appears to be related to the 
source state or its position in the CD.  In a {\it Ginga} 
(1.5--38~keV energy range) observation of GX~5--1 a hard excess was 
detected (Asai et al. 1994).\footnote{Note, however, that the detection 
of a hard tail from GX~5--1 was not confirmed by SIGMA observations 
(Barret \& Vedrenne 1994). }
This component was
fit by a power law with photon index 1.8 (Asai et al. 1994), and 
its intensity decreased from the NB to 
the FB, {\it i.e.} from low to high mass accretion rate.
In a BeppoSAX observation of GX~17+2 the hard component 
(power-law photon index of $\sim 2.7$) was detected in the HB 
and its intensity significantly decreased in the NB (Di Salvo et al. 
2000). Cir X--1, thought to be a peculiar Z source (Shirey et al. 1998), 
was observed by BeppoSAX in the FB (Iaria et al. 2001).  Also
in this case a hard tail was detected in the 
non-flaring spectrum.  
We note that hard component detected here in GX~349+2 is one of the hardest
among the high energy components detected so far in bright LMXBs. 
It corresponds to a photon index of $\sim 1.9$, with no evidence of a high 
energy cutoff in the BeppoSAX range. In fact, using thermal models to fit 
it, we obtain electron temperatures $\ga 100$~keV. 
The power-law tail observed in the NB of GX~5--1 
(Asai et al. 1994), had a photon index of $\sim 1.8$, similar to that 
we measured for GX~349+2. 
Also, Cir X--1 showed a spectral state similar to that of GX~349+2
observed here (Iaria et al. 2001). However, in Cir X--1 the power law was 
much steeper (photon index $\sim 3.3$) than in the case
of GX~349+2, and the low energy continuum was characterized by a softer
electron temperature of the Comptonized component $\la 1$~keV.
In all these cases the hard component seems to become weaker for higher 
accretion rates (see also Fig.~5), while the relative contribution of the 
hard component to the flux in a given source state appears to be different 
in different sources.  Yet, in recent HEXTE observations of
Sco X--1, a hard power-law tail was detected
in 5 out of 16 observations, without any clear correlation with the 
position in the CD (D'Amico et al. 2000). 

It is already known that some Atoll sources, the so-called X-ray bursters,
can show a hard component in their spectra, similar to the observed spectra 
of accreting Black Holes (BH).
Given that BHs and NSs are hardly distinguishable by their broad band spectral
shape (see also Barret et al. 2000, and references therein), 
Barret et al. (1996) introduced another criterion
based on the comparison of hard and soft luminosities. 
Plotting the 1--20 keV luminosity versus the 20--200 keV
luminosity for BHs and NSs of the Atoll class, they observed that all NS 
systems in which a hard component had been detected lie in the so-called 
X-ray burster box, while all BH systems lie outside (see Fig.~5).
If we plot in the same diagram the luminosities of the Z sources GX~17+2 
(Di Salvo et al. 2000), Cir~X--1 (Iaria et al. 2001), and GX~349+2
we find that these data lie clearly outside the X-ray burster box, and
there is no clear distinction between Z sources and BHs (Fig.~5). 
Stated differently, this result shows that BHs are not the only sources
to posses high energy tails when their 1--20 keV luminosity exceeds
$\sim 1.5 \times 10^{37}$ ergs/s.  In particular, the hard components in 
Z sources seem to be similar to the extended power-law tails detected in 
the so-called very-high state (and perhaps intermediate state) of Galactic 
BH systems (e.g. Grove et al. 1998).  On the other hand it appears to be 
still true that only BHs can emit bright hard X-ray tails,
with a 20--200 keV luminosity $\ga 1.5 \times 10^{37}$ erg/s. This 
might be related to the higher Eddington luminosity that on average 
characterizes BHs.
The fact that the luminosity in the hard tail observed in Z sources is 
similar in terms of Eddington luminosity to that seen in BHs
is a further evidence that probably the same mechanism originates 
the hard tails in both BHs and NSs. This would imply that this 
mechanism does not depend on the presence or absence of a solid surface.
\begin{figure}[h!]
\centerline
{\psfig
{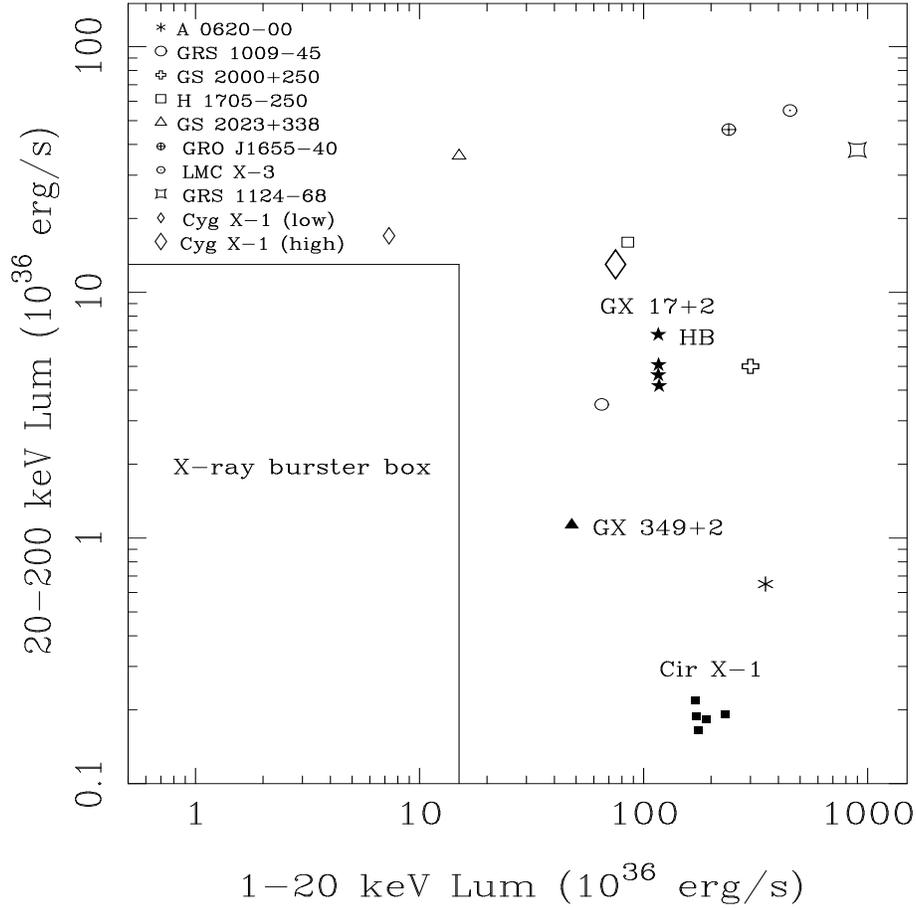}}
\caption{20-200 keV versus 1-20 keV luminosities of BH binaries 
(open symbols, from Barret et al. 2000) and NS type-Z binaries (filled 
symbols).  The so-called {\it X-ray burster box} is plotted as a
solid line. Its boundaries are defined as in Barret et al. (2000). }
\label{fig5}
\end{figure}

As in BHs, the hard tails observed in Z sources can be produced either in a
thermal or non-thermal corona (e.g. Poutanen \& Coppi 1998) or in a bulk
motion of matter close to the NS (e.g. Titarchuk \& Zannias 1998;
Papathanassiou \& Psaltis 2001).  Fast
radial converging motions are unlikely to be dominant in the innermost region 
of the accretion flow in such high-luminosity systems, because of the strong 
radiation pressure emitted from or close to the NS surface.
However, power-law tails, dominating the spectra at high energy, can
also be produced when the flows are mildly relativistic ($v/c \sim 0.1$)
or when the velocity field does not converge (Psaltis 2001).  Therefore 
azimuthal motions around the NS or outflows can be the probable origin of
these components, with flatter power laws corresponding to higher optical
depth of the scattering medium and/or higher bulk electrons velocities,
in a way that is similar to thermal Comptonization (see Psaltis 2001). 
It has been proposed that non-thermal, high energy electrons, 
responsible for the hard tails observed in Z sources, might originate in a jet 
(Di Salvo et al. 2000, Iaria et al. 2001; see also Fender 2001 for a 
review regarding both BH and NS systems).
In fact all the Z sources are detected as variable radio sources, with 
the highest radio fluxes associated with the HB. The radio 
emission weakens in the NB, and is not detected any longer  
in the FB (Hjellming \& Han 1995, Fender \& Hendry 2000, and 
references therein). 
This hypothesis is in agreement with the behavior of GX~17+2, where
the hard tail was observed in the HB. In GX~349+2 the hard
tail is present at the NB/FB vertex, where the radio flux should be
near its minimum, and probably weakens when the source moves further
into the FB.  However, note that a similar correlation was not observed 
in the case of Sco X--1 (D'Amico et al. 2000).
Further observations will help clarifying the
correlation between hard X-ray and radio emission in these sources.

\acknowledgments
The authors thank D. Barret and L. Titarchuk for interesting discussions.  
This work was supported by the Italian Space Agency (ASI), by the Ministero
della Ricerca Scientifica e Tecnologica (MURST).

\clearpage

\section*{TABLES}

\begin{table}[h!]
\small
\begin{center}
\caption{Results of the fitting of the GX~349+2 spectra in the 0.12--200~keV 
energy band.  
The model consists of a blackbody, a Comptonized spectrum 
modeled by {\tt comptt}, a power law, two Gaussian emission lines, and an
absorption edge.
Uncertainties are 90\% confidence level for a single parameter of interest.
The power-law normalization is in units of ph~keV$^{-1}$ cm$^{-2}$ s$^{-1}$ at 
1~keV. The total unabsorbed flux is in the 0.1--200~keV energy range.
The effective exposure time was a factor of $\sim 20$ larger in the 
non-flaring spectrum than in the flaring.}
\label{tab1}
\begin{tabular}{l|c|c} 
\tableline \tableline
 Parameter       & Non-Flaring & Flare \\
\tableline

$N_{\rm H}$ $\rm (\times 10^{22}\;cm^{-2})$
& $0.66^{+0.03}_{-0.02}$  & $0.64 \pm 0.3$ \\

$k T_{\rm BB}$ (keV)
& $0.51 \pm 0.01$ & $0.59 \pm 0.02$ \\

R$_{\rm BB}$ (km)
& $36 \pm 2$ & $31 \pm 2$ \\ 

$k T_{\rm W}$ (keV)
& $1.03 \pm 0.03$ & $1.37 \pm 0.04$ \\

$k T_{\rm e}$ (keV)
& $2.65 \pm 0.05$ & $2.95 \pm 0.07$ \\

$\tau$
& $11.7 \pm 0.4$ & $10.5 \pm 0.5$ \\

R$_{\rm W}$ (km)
& $8.8 \pm 0.6$ & $7.0 \pm 0.5$ \\

Photon Index 
& $1.9^{+0.4}_{-0.3}$  &  1.9 (fixed) \\

Power-law N 
& $3^{+9}_{-2} \times 10^{-2}$  &  
$<2.3 \times 10^{-2}$ \\

$E_{\rm Fe}$ (keV)
& $6.73 \pm 0.05$ & 6.73 (fixed) \\

$\sigma_{\rm Fe}$ (keV)
& $0.31 \pm 0.08$ & 0.31 (fixed) \\

I$_{\rm Fe}$ $(\times 10^{-3}$ ph cm$^{-2}$ s$^{-1})$
& $8.8 \pm 1.5$ & $8.5 \pm 3.7$ \\

Fe Eq. Width (eV)
&  71 & 34  \\

$E_{\rm LE}$ (keV)
& $1.16 \pm 0.03$ & -- \\

$\sigma_{\rm LE}$ (keV)
& $0.05 \pm 0.05$ & -- \\

I$_{\rm LE}$ $(\times 10^{-2}$ ph cm$^{-2}$ s$^{-1})$
& $1.9^{+0.8}_{-0.5}$ & -- \\

LE Eq. Width (eV)
&  17 & --  \\

$E_{\rm edge}$ (keV)
& $8.9 \pm 0.2$ & --   \\

$\tau_{\rm edge}$ ($\times 10^{-2}$)
& $4 \pm 1$ & --  \\

Flux (ergs cm$^{-2}$ s$^{-1}$)
& $1.91 \times 10^{-8}$ & $3.26 \times 10^{-8}$ \\

$\chi^2_{red}$ (d.o.f.)
& 1.17 (189) & 1.22 (119) \\

\tableline
\end{tabular}
\end{center}
\end{table}


\clearpage

\begin{thebibliography}{}
\bibitem[]{}
Agrawal, P. C., et al., 1971, Ap\&SS, 10, 500
\bibitem[]{}
Asai, K., et al., 1994, PASJ, 46, 479
\bibitem[]{}
Barret, D., McClintock, J. E., Grindlay, J. E., 1996, ApJ, 473, 963
\bibitem[]{}
Barret, D., Olive, J. F., Boirin, L., Done, C., Skinner, G. K., Grindlay, 
J. E., 2000, ApJ, 533, 329
\bibitem[]{}
Barret, D., \& Vedrenne, G., 1994, ApJS, 92, 505
\bibitem[]{}
Boella, G., Butler, R. C., Perola, G. C., Piro, L., Scarsi, L., Blecker, J.,
1997a, A\&AS, 122, 299
\bibitem[]{}
Boella, G., et al., 1997b, A\&AS, 122, 327
\bibitem[]{}
Buselli, G., Clancy, M. C., Davison, P. J. N., Edwards, P. J.,
McCracken, K. C., Thomas, R. M., 1968, Nature, 219, 1124
\bibitem[]{}
Christian, D. J., \& Swank, J. H., 1997, ApJS, 109, 177
\bibitem[]{}
Cooke, B. A., Ponman, T. J., 1991, A\&A, 244, 358
\bibitem[]{}
D'Amico, F., Heindl, W. A., Rothschild, R. E., Gruber, D. E., ApJL, submitted
(astro-ph/0008279)
\bibitem[]{}
Di Salvo, T., et al., 2000, ApJL, 544, in press
\bibitem[]{}
Fender, R. P., 2001, Proc. International Symposium on High Energy Gamma-Ray 
Astronomy, Heidelberg, Eds. F. Aharonian \& H. Voelk, AIP, in press
(astro-ph/0101233)
\bibitem[]{}
Fender, R. P., \& Hendry, M. A., 2000, MNRAS, 317, 1
\bibitem[]{}
Frontera, F., et al., 1997, A\&AS, 122, 357
\bibitem[]{}
Greenhill, J. G., Coe, M. J., Burnell, S. J. B., Strong, K. T., 
Carpenter, G. F., 1979, MNRAS, 189, 563
\bibitem[]{}
Grove, J. E., Johnson, W. N., Kroeger, R. A., McNaron-Brown, K., 
Skibo, J. G., Phlips, B. F., 1998, ApJ, 500, 899
\bibitem[]{}
Hasinger, G., \& van der Klis, M., 1989, A\&A, 225, 79
\bibitem[]{}
Hasinger, G., van der Klis, M., Ebisawa, K., Dotani, T., Mitsuda, K.,
1990, A\&A, 235, 131
\bibitem[]{}
Haymes, R. C., Harnden, F. R., Johnson, W. N., Prichard, H. M., Bosch, H. E.,
1972, ApJ, 172, L47
\bibitem[]{}
Hjellming, R. M., \& Han, X. H., 1995, in {\it X-ray Binaries}, , Lewin
W. H. G., van Paradijs J., van den Heuvel E. P. J. eds., Cambridge
Astrophysics Series, 308
\bibitem[]{}
Iaria, R., Burderi, L., Di Salvo, T., La Barbera, A., Robba, N. R., 2001,
ApJ, 547, in press (astro-ph/0009183)
\bibitem[]{}
In't Zand, J. J. M., et al., 1999, A\&A, 345, 100
\bibitem[]{}
Jain, A., et al., 1984, A\&A, 140, 179
\bibitem[]{}
Kallman, T. R., Liedahl, D., Osterheld, A., Goldstein, W., Kahn, S., 
1996, ApJ, 465, 994
\bibitem[]{}
Levine, A. M., et al., 1984, ApJS, 54, 581
\bibitem[]{}
Magdziarz, P., \& Zdziarski, A. A., 1995, MNRAS, 273, 837
\bibitem[]{}
Manzo, G., Giarrusso, S., Santangelo, A., Ciralli, F., Fazio, G., Piraino, 
S., Segreto, A., 1997, A\&AS, 122, 341
Masetti, N., et al., 2000, A\&A, in press (astro-ph/0009044)
\bibitem[]{}
Mitsuda, K., et al., 1984, PASJ, 36, 741
\bibitem[]{}
Miyamoto, S., \& Matsuoka, M., 1977, SSRv, 20, 687
\bibitem[]{}
Nishimura, J., Mitsuda, K., Itoh, M., 1986, PASJ, 38, 819
\bibitem[]{}
Papathanassiou, H., \& Psaltis, D., 2001, MNRAS, in press (astro-ph/0011447) 
\bibitem[]{}
Parmar, A. N., et al., 1997, A\&AS, 122, 309
\bibitem[]{}
Peterson, L. E., \& Jacobson, A. S., 1966, ApJ, 145, 962
\bibitem[]{}
Poutanen, J., \& Coppi, P. S., 1998, Phys. Scripta, T77, 57
\bibitem[]{}
Psaltis, D., 2001, ApJ, in press (astro-ph/0011534)
\bibitem[]{}
Riegler, G. R., Boldt, E., \& Serlemitsos, P., 1970, Nature, 266, 1041 
\bibitem[]{}
Shirey, R. E., Bradt, H. V., Levine, A. M., Morgan, E. H., 1998, ApJ, 506, 374
\bibitem[]{}
Soong, Y., Rothschild, R. E., 1983, ApJ, 274, 327
\bibitem[]{}
Strickman, M., \& Barret, D., Detections of multiple hard X-ray 
flares from Sco X-1 with OSSE, in AIP Conf. Proc. 510, Proc.  of the 
Fifth Compton Symposium, Eds M.L. McConnel and J.M. Ryan 
(New York:AIP), 222-226, 2000
\bibitem[]{}
Sunyaev, R. A., Titarchuk, L. G., 1980, A\&A, 86, 121
\bibitem[]{}
Titarchuk, L., 1994, ApJ, 434, 570
\bibitem[]{}
Titarchuk, L., \& Zannias, T., 1998, ApJ, 493, 863
\bibitem[]{}
Ubertini, P., Bazzano, A., Cocchi, M., La Padula, C., Sood, R. K.,
1992, ApJ, 386, 710
\bibitem[]{}
Valinia, A., \& Marshall, F. E., 1998, ApJ, 505, 134 
\bibitem[]{}
Zhang, W., Strohmayer, T. E., Swank, J. H., 1998, ApJ, 500, L167
\end{thebibliography}
\end{document}